\documentclass[aps,prl,preprint,groupedaddress]{revtex4}
\usepackage{graphicx}
\usepackage[english]{babel}

\begin{document}

\title{Distinguishing Cause from Correlation in Tokamak Experiments to Trigger
  Edge Localised Plasma Instabilities}

\author{\fontshape{sc}\selectfont{ Anthony J. Webster}$^{1,2}$}
\author{JET EFDA Contributors\footnote{See the Appendix of
    F. Romanelli et al., Proceedings of the 24th IAEA Fusion Energy
    Conference 2012, San Diego, US.}}
\affiliation{$^1$ JET-EFDA, 
  Culham Science 
  Centre, Abingdon, OX14 3DB, UK}
\affiliation{$^2$ CCFE, 
  Culham Science 
  Centre, Abingdon, OX14 3DB, UK}


\date{\today}

\begin{abstract}

The generic question is considered: How can we determine the
probability of an otherwise quasi-random event, having been triggered
by an external influence?  
A specific problem is the quantification of the success of techniques
to trigger, and hence control, edge-localised plasma instabilities
(ELMs) in magnetically confined fusion (MCF) experiments. 
The development of such techniques is essential to ensure tolerable
heat loads on components in large MCF fusion devices, and is necessary
for their development into economically successful power plants. 
Bayesian probability theory is used to rigorously formulate the
problem and to provide a formal solution.    
Accurate but pragmatic methods are developed to estimate triggering 
probabilities, and are illustrated with experimental data.   
These allow results from  experiments to be quantitatively assessed,
and rigorously quantified conclusions to be formed.     
Example applications include assessing whether triggering of ELMs
is a statistical or deterministic process, and the establishment of
thresholds to ensure that ELMs are reliably triggered. 
\end{abstract}


\maketitle

\section{Introduction}

\begin{figure}[htbp!]
\begin{center}
\includegraphics[width=\columnwidth,height=10cm]{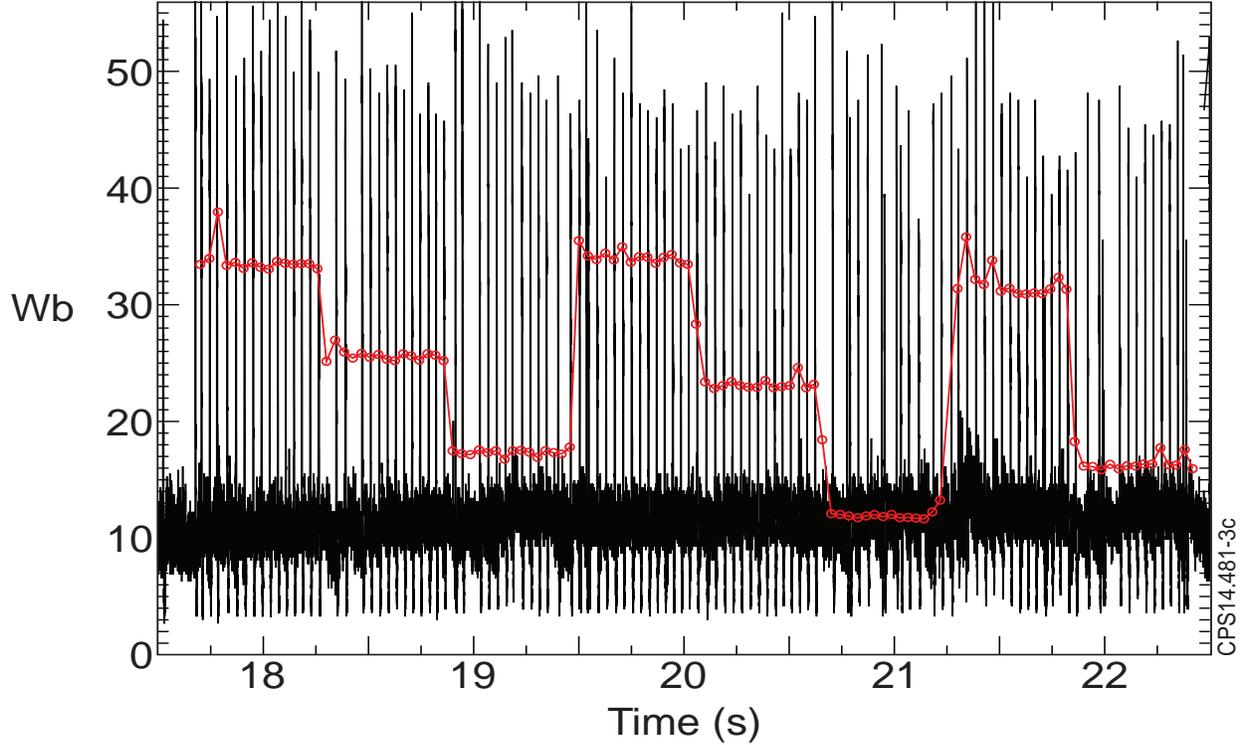}
\vspace{.0cm}
\caption{\label{Webers} JET plasma 83440. 
  The measured Volt seconds (Wb) of each kick is
  plotted (red circles) versus the time in seconds during the pulse,
  with the strong Be II light emissions due to 
  ELMs superimposed (not to scale). 
  For kicks with large amplitude in Wb, ELMs appear to be clearly
  synchronised with the kicks, but for small kicks such as
  those between 20.6s and 21.4s their influence is much less clear. 
}
\end{center}
\end{figure}

Presently the most successful 
high-performance nuclear fusion experiments use a toroidal magnetic
field to confine the plasma in a tokamak \cite{Wesson}. 
Unfortunately the most highly confined of these plasmas are
susceptible to edge-localised instabilities (``ELMs'') \cite{Zohm},
that eject particles and energy onto the plasma-facing components. 
In future power-plant scale devices, the size of the ELMs must be
limited to avoid components being damaged. 
One proposal is to deliberately and rapidly trigger ELMs, with the hope
of having a larger number of smaller ELMs. 
A method that has been found successful at triggering ELMs is to apply
a rapidly increasing radial magnetic field to the plasma, that
``kicks'' the plasma vertically to trigger ELMs
\cite{Degeling,Lang,Liang,LangPacing}.  
To understand and optimise the method it is necessary to explore the
threshold between when the kicks trigger ELMs, and when they simply
perturb the plasma. 
For example figure \ref{Webers} shows a scan of kick sizes, with a
systematic variation of the
amplitude and duration of the voltage applied to coils that produce the
radial magnetic field that can trigger the ELMs. 
For these situations it is necessary to be able to quantify the
success by which ELMs are triggered, so as to better understand
physically what is triggering the ELM, and practically, what is
required to successfully do so. 
A complication is that ELMs occur regularly without kicks, so how can
we determine whether an ELM is due to a kick or is naturally occurring? 
A major step towards this is to understand the time-scales over which
the kick influences the plasma most strongly. 
This then determines a narrower time window when a kick is able to
trigger an ELM, but it does not rule out the possibility that some
ELMs occur naturally within that time window also. 
Correctly accounting for this possibility is a purpose of this
paper. 
In the next section we outline how Bayesian probability theory can be
used to formulate a general solution to this problem. 
The generic problem however is much broader - for example, how can you
tell if the increase in insurance claims due to lightning strikes are
due to global warming or statistical chance? 
It might be possible to adapt some of the methods here to those
broader questions. 
The following sections explore how the solution can be applied to 
the specific question of determining the success of
``kick-triggering'' experiments.


\section{Formulation and formal solution of the problem}

Bayesian probability theory starts 
from some simple but reasonable requirements about a theory
of probability, and shows how the usual laws of probability theory can
be derived from them \cite{Jaynes}, such as the basic product rule, 
\begin{equation}\label{e1}
P(A,B|C) = P(A|B,C)P(B|C) = P(B|A,C)P(A|C)
\end{equation}
and the basic sum rule,
\begin{equation}\label{e2}
P(A|C) + P(\bar{A}|C) = 1 
\end{equation}
where $P(A,B|C)$ reads as the probability of $A$ and $B$ being
simultaneously true given that $C$ is true, 
and $P(A|B,C)$ reads as the probability of $A$ 
being true given that $B$ and $C$ are true, 
and $\bar{A}$ is used to denote the
negation of $A$ (i.e. ``not $A$''). 
These relations can be used to formulate a variety of 
mathematical expressions that relate the probability of events
\cite{Jaynes}. 
One useful expression that is derived in the appendix for convenience
is,  
\begin{equation}\label{e3}
P(A|C) = P(A|B,C) P(B|C) + P(A | \bar{B},C) P(\bar{B}|C) 
\end{equation}
In the context of our subsequent discussion this will be used to
obtain the probability of the event $A$ occurring at time $t$, in
terms of the probability of event $A$ occurring at time $t$ given that
action $B$ has (or has not) triggered the event, with $C$
determining the time at which the triggering is attempted. 
Specifically we consider the probability density of observing the next
ELM at time $t$ to $t+dt$ after a ``vertical kick'' starts to
influence the plasma at $t=\tau_m$, with, 
\begin{equation}\label{p0}
P(t|\tau_m) = P(t|K,\tau_m) P(K|\tau_m) + P(t|\bar{K},\tau_m)
P(\bar{K}|\tau_m) 
\end{equation}
where $P(t|\tau_m)$ is the probability density of observing an ELM at
time $t$ after a (vertical) kick starts to influence the plasma at
time $\tau_m$,  
$P(t|K,\tau_m)$ is the probability density of an ELM at time $t$ after
$\tau_m$ given that a kick at has successfully triggered an ELM, 
and $P(t|\bar{K},\tau_m)$ is the probability density of an ELM at time
$t$ after $\tau_m$ if a kick has failed to trigger an ELM. 
$P(K|\tau_m)$ is the probability of a kick 
successfully triggering an ELM and $P(\bar{K}|\tau_m)$ is the
probability of a kick  
not triggering an ELM, given that the kick starts to
influence the plasma at time $\tau_m$.   
Because of (\ref{e2}), $P(\bar{K}|\tau_m)=1 - P(K|\tau_m)$, and we
have,  
\begin{equation}\label{p1}
P(t|\tau_m) = P(t|K,\tau_m) P(K|\tau_m) + P(t|\bar{K},\tau_m) ( 1 -
P(K|\tau_m) ) 
\end{equation}
This equation is exact, and can be developed in various ways. 
The aim is to construct a mathematical expression whose terms can be
accurately estimated from the available experimental data, allowing
$P(K|\tau_m)$ to be determined. 
Here (\ref{p1}) will be used to formalise the idea that triggered ELMs will
appear in a short time interval after a kick, using the
information to rigorously define the probability of an ELM having
been triggered. 
Crucially, this is done in a way that allows the possibility that some of the
observed ELMs will have occurred naturally, the resulting estimates 
quantitatively account for this.

\begin{figure}[htbp!]
\begin{center}
\includegraphics[width=\columnwidth,height=10cm]{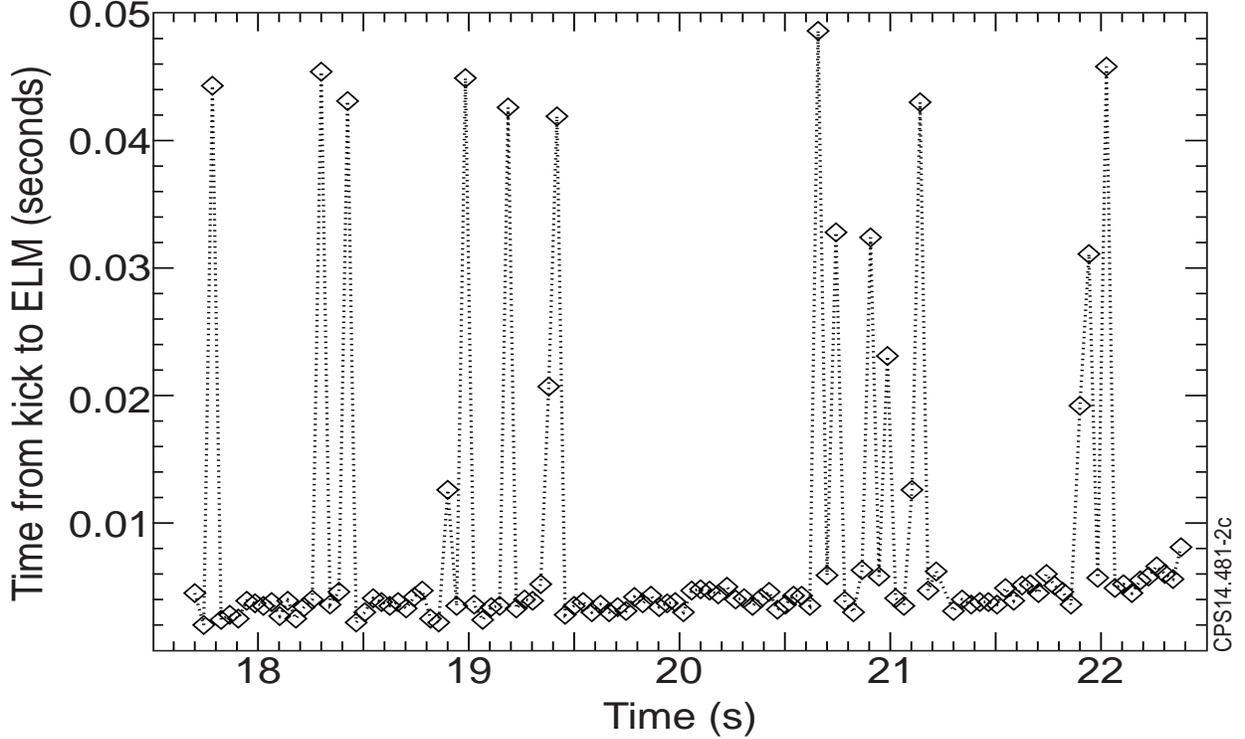}
\vspace{.0cm}
\caption{\label{KickDetect} 
  JET plasma 83440. 
  The time from the start of a kick to the next ELM is plotted versus
  time during the plasma pulse. 
  Due to the limited kick duration and the time for the kicked
  magnetic field to diffuse through the plasma, kicks can only
  influence  the plasma for of order 1-8 ms after they start. 
  Most of the ELMs bunched within this time interval will have been
  triggered by a kick, although some may have occurred naturally. 
}
\end{center}
\end{figure}

Some background. 
Following a ``kick'' in the JET tokamak, there is a time period of order
$1$-$1.5$ milliseconds before the kick's magnetic field starts to
strongly interact with the plasma, and another time period
proportional to the kick's duration, before the kick reaches its
maximum strength. 
The time delay is due to the time needed for the magnetic field that
 the control coils are trying to produce, to diffuse through
JET's metal vacuum vessel and other conducting components.
By observing ELMs that are clearly triggered and paced at the
kick-frequency, it appears that triggered ELMs occur approximately
within $\tau_m$ to $\tau_m+ \Delta \tau$, where $\Delta \tau$ is the
kick's duration. 
For example see figure \ref{KickDetect}, whose results are consistent
with a resistive diffusion of the ``kicked'' magnetic field through
JET's metal vacuum vessel over a 1.0-1.5 ms timescale, and ELMs being
triggered after 
the magnetic perturbation has grown towards its maximum value over a
timescale of 1.5-4.5ms 
depending on the kick's duration and amplitude. 
However, we cannot infer the kick success simply by counting the
fraction of ELMs that occur in this time period after a kick, because a
fraction of those will be expected to occur in that time interval
naturally.  
Eq. \ref{p1} accounts for this through the term
$P(t|\bar{K},\tau_m)P(\bar{K}|\tau_m)$, 
the probability of observing ELMs at a time
$t$ since a kick, given that the ELM is not triggered. 
Formally all of this can be achieved by integrating both sides of
(\ref{p1}) from 
$\tau_m$ to $\tau_m+\Delta \tau$, with respect to $t$.  
Then after integration we get, 
\begin{equation}
P_{\Delta \tau} = P(K) + P_{\bar{K}} \left( 1 - P(K) \right) 
\end{equation}
where 
$P_{\Delta \tau} \equiv  \int_{\tau_m}^{\tau_m+\Delta \tau} P(t) dt$, 
$\int_{\tau_m}^{\tau_m+\Delta \tau} P(t|K)
dt= 1 $ by the definition that these kicks will successfully trigger
ELMs, and 
$P_{\bar{K}} \equiv  \int_{\tau_m}^{\tau_m+\Delta \tau} P(t|\bar{K})
dt$ is the probability of ELMs occurring 
naturally and being observed within $\tau_m$ to
$\tau_m+\Delta \tau$ without having been triggered by kicks.  
Consequently, 
\begin{equation}\label{PK}
P(K) = \frac{P_{\Delta \tau}-P_{\bar{K}}}{1 - P_{\bar{K}}} 
\end{equation}
$P_{\Delta \tau}$ and $P_{\bar{K}}$ both depend on $\Delta \tau$ and
$\tau_m$, although for reasonably large numbers of kicks we will find
that $P_{\bar{K}}$ (and consequently $P_{\Delta \tau}$), are
approximately independent of $\tau_m$. 
Notice that if we had ignored naturally occurring ELMs, with
$P_{\bar{K}}=0$, we would simply have $P(K) = P_{\Delta \tau}$, 
or in other words, we would estimate the probability of a kick
successfully triggering an ELM simply as the probability of
observing an ELM in the ``kick-triggered'' time interval. 
If $P_{\bar{K}}$ is sufficiently small then the estimates of
$P(K)=P_{\Delta \tau}$ and (\ref{PK}) will be similar. 
Also if $P_{\bar{K}}$ is small, then, 
\begin{equation}\label{APK}
P(K) = P_{\Delta \tau} - P_{\bar{K}} \left( 1 - P_{\Delta \tau}
\right)  + O \left( P_{\bar{K}}^2 \right)
\end{equation}
which is useful for obtaining simple leading-order error estimates as
the sum of errors in $P_{\Delta \tau}$ and $P_{\bar{K}}$, which are
easy to calculate and record. 
For kick-triggering studies, $P_{\bar{K}}$ is usually small. 
$P_{\Delta \tau}$ can be estimated from the kick-ELM data by counting
the number of ELMs that occur naturally within the set of time
intervals $\tau_m$ to $\tau_m+\Delta \tau$; this will be considered in
a later section. 
$P_{\bar{K}}$ can be estimated from equivalent ELMing plasma data
without kicks, which we discuss next.

\section{Estimating $P_{\bar{K}}$}

Next we consider how to estimate $P_{\bar{K}}$ from ELM time data that
does not have kicks or other triggering mechanisms. 
Firstly we calculate the probability of an ELM being observed in a
time interval $\tau_m$ to $\tau_m +\Delta \tau$, finding a simple
estimate for the limit where $\tau_m \sim n \bar{t}$ has $n\gg 1$. 
This is a useful approximation, so we explore when it is reasonable,
and test the estimate by calculating the number of
ELMs that we would expect to occur naturally in the time intervals
($\tau_m$,$\tau_m+\Delta \tau$) that we might incorrectly think are
``triggered'' ELMs, and compare this estimate with observations from
experimental data.

The probability of observing the $n$th ELM in ($\tau_m$,$\tau_m +
\tau$), is, 
\begin{equation}\label{Psum1}
\int_{\tau_m}^{\tau_m + \tau} p(x|n) dx 
\end{equation}
where $x=\sum_{i=1}^n t_i$, and $t_i$ are the waiting times between
the $(i-1)$th and $i$th ELM. 
The probability of {\sl any} ELM being in ($\tau_m$, $\tau_m + \tau$),
is, 
\begin{equation}\label{pkbar0}
P_{\bar{K}}(\tau_m) = 
\sum_{n=1}^{\infty} \int_{\tau_m}^{\tau_m + \tau} p(x|n) dx
\end{equation}
for the present we have decided to explicitly emphasise the dependence
of $P_{\bar{K}}$ on $\tau_m$ by writing $P_{\bar{K}} (\tau_m)$.  
As is outlined in 7.16 of Jaynes \cite{Jaynes}, $p(x|n)$ can be
written as, 
\begin{equation}\label{pxnft}
p(x|n) = \frac{1}{2\pi} \int_{-\infty}^{\infty} \psi (\alpha)^n
e^{-i\alpha x} d\alpha 
\end{equation}
with,
\begin{equation}\label{psialpha}
\psi(\alpha) = \int_{-\infty}^{\infty} P(t) e^{i\alpha t} dt 
\end{equation}
where $P(t)$ is the ELM waiting time pdf, that gives the probability
of observing an ELM in time $t$ to $t+dt$ as $P(t)dt$. 
For many situations $P(t)$ can be reasonably approximated by a
Gaussian \cite{WebsterDendy2013}, and for those cases we exactly have, 
\begin{equation}\label{CLT}
p(x|n) = \frac{1}{\sqrt{2\pi n \sigma^2}} 
\exp\left( -\frac{(x-n\bar{t})^2}{2n\sigma^2} \right)
\end{equation}
where $\bar{t}$ is the average ELM waiting time and $\sigma$ is the
ELM waiting times' standard deviation. 
These two quantities $\bar{t}$ and $\sigma$ 
can be estimated from the ELM time data with an error of
order $1/\sqrt{M}$, where $M$ is the total number of observed ELMs
\cite{Sivia}. 
Equation (\ref{CLT}) is exact \cite{Jaynes} whenever $P(t)$
is a Gaussian, 
however if $n \gg 1$ then with
only a small number of 
exceptions \cite{Jaynes}, the Central Limit Theorem ensures that
(\ref{CLT}) remains true independent of the 
specific form of ELM waiting time pdf \cite{Jaynes}. 
Using (\ref{pkbar0}) and (\ref{CLT}) the probability of an ELM in time
$\tau_m$ to $\tau_m + \tau$ is,  
\begin{equation}
P_{\bar{K}} (\tau_m) = \sum_{n=1}^{\infty} \int_{\tau_m}^{\tau_m + \tau} 
\frac{dx}{\sqrt{2\pi n \sigma^2}} 
\exp\left( -\frac{(x-n\bar{t})^2}{2n\sigma^2} \right)
\end{equation}
For $n\bar{t} \gg \bar{t}$ ($n\gg 1$) , the sum can be approximated by
an integral, with $y=n\bar{t}$ and $dy=\bar{t}$, giving, 
\begin{equation}\label{Pint}
P_{\bar{K}} = \int_{0}^{\infty} \frac{dy}{\bar{t}} 
\int_{\tau_m}^{\tau_m + \tau} 
\frac{dx}{\sqrt{2\pi y (\sigma^2/\bar{t})} }
\exp\left( -\frac{(x-y)^2}{2y(\sigma^2/\bar{t})} \right)
\end{equation}
where the dependence of $P_{\bar{K}}$ on $\tau_m$ has been removed
for reasons that will be explained shortly. 
Changing variables, with $x=\hat{x} \sigma^2/\bar{t}$ and $y=v^2
\sigma^2/\bar{t}$, so that $ dy/\sqrt{y} = 2 dv (\sigma^2/\bar{t})$,
and exchanging the order of integration, gives, 
\begin{equation}
P_{\bar{K}} =
\frac{1}{\bar{t}} 
\int_{\tau_m(\sigma^2/\bar{t})}^{(\tau_m + \tau)(\sigma^2/\bar{t})}
d\hat{x} 
\left(\frac{\sigma^2}{\bar{t}} \right)
\frac{1}{\sqrt{2\pi}}  
 \int_{0}^{\infty} dv 
\exp\left( \frac{(x-v^2)^2}{2v^2} \right)
\end{equation}
Using the integral \cite{Schulman}, 
\begin{equation}
\int_0^{\infty} dx \exp \left( -a x^2 - \frac{b}{x^2} \right) 
= \sqrt{ \frac{\pi}{4a} } \exp \left( -2 \sqrt{ab} \right) 
\end{equation}
it can easily be shown that, 
\begin{equation}
\int_0^{\infty} dv \exp \left( -\frac{(\hat{x}-v^2)^2}{2v^2} \right) 
= \sqrt{ \frac{\pi}{2} }
\end{equation}
a surprising result, because like a normalised probability
distribution with a parameter $\hat{x}$, the result is independent of
$\hat{x}$.  
Consequently, doing the final integral over $\hat{x}$ and cancelling
terms,  we get, 
\begin{equation}\label{PDT}
P_{\bar{K}} = \frac{\Delta \tau}{\bar{t}}
\end{equation}

A surprising and disappointing aspect of (\ref{PDT}) is that $\tau_m$
does not appear, 
and the influence of correlations when $\tau_m \simeq \bar{t}$ for
example, are not captured by the calculation. 
The only approximation that is always present 
enters subtly, through the need of $n\gg 1$ for the sum to be
accurately approximated by an integral. 
The approximation $n\gg 1$ removes the information about the discrete
nature of ELMs and consequently is unable to describe the influence of
correlations between $\tau_m$ and $\bar{t}$. 
Consider the example with $\tau_m$ small. 
Then (\ref{Psum1}) could be accurately approximated by keeping only the
first few terms in the sum over $n$, and the approximation of
(\ref{Pint}) would be poor. 
The influence of correlations between the natural ELM frequency and
the kick frequency is discussed in the Appendix, where it finds that
provided $\Delta \tau/\sigma$ is reasonably small, where $\sigma$ is
the standard deviation in the natural ELM period, then any coherence
between the natural ELMs and 
kick frequency will be rapidly lost within the first one or possibly
two ELMs. 
Another caveat is that if
$P(t)$ is non-Gaussian, then $n \gg 1$ is required for the central
limit theorem to ensure that $p(x|n)$ will tend to (\ref{CLT}); but
occasionally for some distributions even if $n\gg 1$ the rate of
convergence can be slow and the estimates less good than expected. 
In summary, for small $\tau_m/\bar{t}$, for which $n\sim 1$ is 
most relevant, then in principle equations (\ref{Psum1}),
(\ref{pxnft}), and (\ref{psialpha}) must be 
used; but despite the caveats mentioned above, 
provided $\tau_m/\bar{t} \gg 1$ and $\Delta \tau/ \sigma$ are 
reasonably small then (\ref{Pint}) and (\ref{PDT}) should in most
cases provide an accurate approximation. 
Estimates for $\bar{t}$ and $\sigma$ must be found from equivalent ELM
data for natural (untriggered) ELMs. 
They can be estimated from a set of $M$ ELM waiting times $\{ t_i \}$,
of $\bar{t}\simeq \frac{1}{M} \sum_{i=1}^M t_i$ 
and $\sigma^2 \simeq \frac{1}{M-1} \sum_{i=1}^M (t_i -\bar{t} )^2$ 
both of which have errors of order $1/\sqrt{M}$ \cite{Sivia}. 
Ideally there will be plenty of data for natural ELMs, so that $M\gg
1$, and we can neglect the errors in $\bar{t}$ and $\sigma$ compared
with the other errors in (\ref{NbarK}). 
Otherwise the leading order errors
give,  
\begin{equation}\label{fpbark}
P_{\bar{K}} \simeq  \frac{\Delta \tau}{\bar{t}} 
\left( 1 
\pm \frac{\sigma}{\bar{t}\sqrt{M}} 
\right)
\end{equation}
with the error estimated from 
$\bar{t} \simeq (1/M)\sum_{i=1}^M t_i \pm \sigma/\sqrt{M}$. 
The greatest source of systematic error is likely to be through
unintended changes to the plasma's properties between pulses, leading
to incorrect estimates for $\bar{t}$ and $\sigma$. 
This risk can be reduced by using a reference pulse from the same
session, or a reference phase from within a pulse, although longer
pulses will give greater statistical accuracy. 
In practice $P_{\bar{K}}$ is comparatively small, and unless the
number of kicks are much larger than the $\sim$15 in the example
considered here, then the errors are dominated by those in
$P_{\Delta\tau}$ that are statistical in origin and can only be
reduced by 
increasing the number of kicks in the study.

\begin{figure}[htbp!]
\begin{center}
\vspace{0.cm}
\includegraphics[width=\columnwidth,height=10cm]{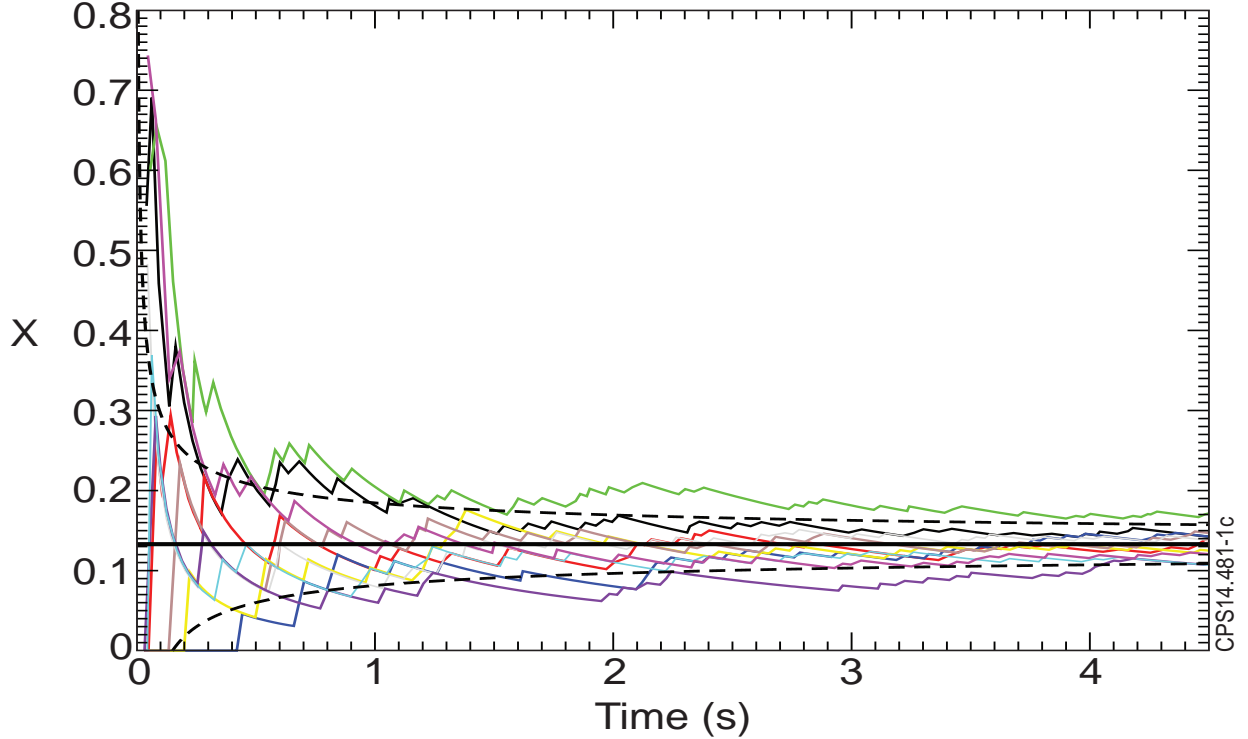}
\vspace{.0cm}
\caption{\label{StatsTest} Comparison between calculated estimates for
  $N_{\bar{K}}$, and observed  from experimental data. 
  For ten equivalent JET plasmas, the number of ELMs are counted that
  occur by chance within the set of intervals $\tau_m$ to $\tau_m +
  \Delta \tau$, that if kicks were present would be presumed to be
  kick-triggered. 
  Theoretical lines are: the estimated average (thick black line), the
  average $\pm$ the standard deviation (dashed black lines). 
  Assuming the data is Gaussianly distribution
  then we would expect 68\% of results to be within the dashed
  lines. 
  The results are reassuringly consistent with this. }
\end{center}
\end{figure}

We can test Eq. \ref{PDT} by calculating the number of natural
(untriggered) ELMs we would expect to observe in the
``kick-triggered'' time intervals, and compare this with the number
that are observed in equivalent time intervals of experimental
data (without kicks).  
Assuming that (\ref{PDT}) is a reasonable 
approximation for $P_{\bar{K}}$, as it very often will be,
then we can estimate $N_{\bar{K}}$ as follows. 
The probability of observing an unordered sequence of $N_{K}$ time
intervals ($\tau_1$,
$\tau_1 +\tau$), ($\tau_2$, $\tau_2 +\tau$), ... , ($\tau_{N_{K}}$, 
$\tau_{N_{K}} + \tau$), with 
$N_{\bar{K}}$ ELMs spread among them is given by, 
\begin{equation}
P(N_{\bar{K}},N_K) = 
\left( 
\begin{array}{c}
N_K
\\
N_{\bar{K}}
\end{array}
\right) 
\left( P_{\bar{K}} \right)^{N_{\bar{K}}}
\left( 1 - P_{\bar{K}} 
\right)^{N_K - N_{\bar{K}}}
\end{equation} 
i.e. the Binomial distribution (because the order in which the
$N_{\bar{K}}$ natural ELMs coincide with an interval
$\tau_i$ to $\tau_i + \Delta \tau$ is not important). 
Consequently the expected number of (untriggered) ELMs
to be observed by chance in the ``triggered ELMs'' time interval can
be calculated from 
$\langle N_{\bar{K}} \rangle = \sum_{N_{\bar{K}=1}}^{N_K} N_{\bar{K}} 
P(N_{\bar{K}},N_K)$, 
and is found using (\ref{PDT}) for $P_{\bar{K}}$ to be,
\begin{equation}\label{ENKb}
\langle N_{\bar{K}} \rangle  = N_K \frac{\Delta \tau}{\bar{t}} 
\end{equation}
The standard deviation is 
$\sqrt{
\langle N_{\bar{K}}^2 \rangle - \langle N_{\bar{K}} \rangle^2 }
= 
\sqrt{N_K \frac{\Delta \tau}{\bar{t}} 
  \left( 1 - \frac{\Delta
      \tau}{\bar{t}} \right)  }$,
which allows an estimate for the error in (\ref{ENKb}). 
Therefore taking $N_{\bar{K}}\simeq \langle  N_{\bar{K}} \rangle$, we
obtain an estimate of,   
\begin{equation}\label{NbarK}
N_{\bar{K}} \simeq N_K \frac{\Delta \tau}{\bar{t}} 
\pm \sqrt{N_K \frac{\Delta \tau}{\bar{t}} } 
\sqrt{1 - \frac{\Delta \tau}{\bar{t}} }
\end{equation}
Eq. \ref{NbarK} allows us to estimate $N_{\bar{K}}$, given the number
of kicks $N_K$, and a reference phase of normal ELMs from which the
ELMs' average waiting time $\bar{t}$ and standard deviation $\sigma$
can be estimated. 
To test the above Eq. \ref{NbarK}, we count the number of ELMs
within time intervals of $n\times$(0.02) to $n\times$(0.02)+0.004
seconds, with $n$= 0,1,2, ... , during the time period of 9-13.5
seconds of ten equivalent JET plasmas that do not have kicks. 
Details of the pulses are in \cite{Brez}, we consider the subset of:
83630, 83629, 83628, 83627, 83626, 83625, 83624, 83640, 83641, and
83642, 
for which the average ELM waiting time across all pulses was $\bar{t}
= 0.030$ seconds.  
Figure \ref{StatsTest} plots the number of ELMs that are counted in
the time intervals that would usually be presumed to be
``kick-triggered'', 
divided by the number of ``kick-triggered'' time intervals (estimated
as the time divided by the period between the intervals), along with
our estimate for $N_{\bar{K}}/N_K$ using Eq. \ref{NbarK}.  
The agreement is surprisingly good. 
If the results were Gaussianly distributed then we would expect
approximately 68\% of the data sets to be within one standard
deviation, indicated on the plots by the dashed lines. 
The results are reassuringly consistent with this.

\section{Estimating $P_{\Delta \tau}$}

The estimation of $P_{\Delta \tau}$ given the observation of $m$ ELMs
in $n$ ``kick-triggered'' time intervals, is a classic and well-known
problem in Bayesian probability theory that we will briefly review
next. 
The probability of observing $m$ successes in $n$ trials, with a
probability $p$ of a success at each trial, is given by the binomial
distribution, 
\begin{equation}\label{B0}
P(m,n|p) = \left(
\begin{array}{c}
n\\m
\end{array}
\right) p^m (1-p)^{n-m} 
\end{equation}
Bayes theorem gives \cite{Sivia,Jaynes}, 
\begin{equation}\label{B1}
P(p|m,n) = \frac{P(m,n|p)P(p)}{\int_0^1 P(m,n|p)P(p) dp} 
\end{equation}
which allows us to obtain a probability distribution for $p$ given the
observed $m$ successes from $n$ trials, and a suitable prior
distribution $P(p)$. 
Present models for ELMs have ELMs being triggered once a threshold of
pressure gradient or current is exceeded \cite{WebsterRev}. 
If we regard kick-triggering in a similar way, with a probability of
success that rapidly steps between zero and one - either a kick is
strong enough to trigger an ELM or it is not, then we might follow
the reasoning of Jaynes (pages 382-386 of Ref. \cite{Jaynes}), that
leads to the ``Haldane'' prior $P(p)$ with, 
\begin{equation}
P(p) = \frac{C}{p(1-p)} 
\end{equation}
with $C$ a constant. 
Using this prior in (\ref{B1}) along with (\ref{B0}) for $P(m,n|p)$,
we get (after doing the integral in the denominator) \cite{Jaynes}, 
\begin{equation}
P(p|m,n)dp = 
\frac{(n-1)!}{(m-1)!(n-m-1)!} 
p^{m-1}(1-p)^{n-m-1} dp 
\end{equation}
This gives an expected value for $p$ of, 
\begin{equation}\label{pbar1}
\langle p \rangle = \frac{m}{n} 
\end{equation}
with a standard deviation $\sigma$ of, 
\begin{equation}\label{Hs1}
\sigma = 
\sqrt{
\frac{m(n-m)}{n^2(n+1)} 
}
\end{equation}

Alternately, we might regard kick-triggering as a statistical process
as in \cite{WebsterDendy2013}, 
with a success probability that increases gradually with the size of
the kick. 
Then it would seem reasonable to take a uniform prior, with $P(p)$ a
constant. 
With this assumption we get, 
\begin{equation}
P(p|m,n)dp = 
\frac{(n+1)!}{m!(n-m)!} 
p^{m}(1-p)^{n-m} dp 
\end{equation}
This gives an expected value for $p$ of, 
\begin{equation}\label{pbar2}
\langle p \rangle = \frac{m+1}{n+2} 
\end{equation}
with a standard deviation $\sigma$ of, 
\begin{equation}\label{Hs2}
\sigma = 
\sqrt{
\frac{(m+1)(n-m+1)}{(n+2)^2(n+3)} 
}
\end{equation}

Whichever prior we regard as being most reasonable, we can 
estimate $P_{\Delta \tau}$ from the $N$ ELMs observed
during the total of $N_K$ kicked time intervals, as, 
\begin{equation}\label{fpdt}
P_{\Delta \tau} \simeq \langle p \rangle \pm \sigma 
\end{equation}
with $\langle p \rangle$ and $\sigma$ estimated from either \ref{pbar1} and
\ref{Hs1}, or \ref{pbar2} and \ref{Hs2}. 
We will evaluate $P_{\Delta \tau}$ with both priors to provide
examples of how the results can become modified.  
Equations \ref{PK}, \ref{fpbark}, and \ref{fpdt} allow us to
rigorously estimate the probability of kicks triggering ELMs, given
the number of kicks $N_K$, a suitable estimate for $\tau_m$ and
$\Delta \tau$, the observed number of ELMs $N$ in the intervals
$\tau_m$ to $\tau_m + \Delta \tau$, and an equivalent phase of plasma
with natural (unkicked) ELMs from which to estimate $N_{\bar{K}}$ from
(\ref{fpbark}). 

\section{Examples: Quantifying kick-triggering success}

A kick consists of a step in the voltage that is applied to the
vertical control coils, that has a duration in time and an amplitude
\cite{KickRef}. 
In a recent JET experiment (pulse 83440), the duration and amplitude
of the kick 
was deliberately varied to explore how kick-triggering success
depended on them. 
To analyse the kick-triggering probability it is necessary to explore
a bit more of the physics involved in a vertical kick.  
The diffusion time of JET's vacuum vessel is of order 1.3ms, so a kick 
of duration $\tau_{kick}$ has negligible influence on the plasma for 1.3ms,
has a maximum perturbation over $\tau_{kick}$ to $\tau_{kick}+1.3$ms, and falls
to less than half of its value in of order $2\times \tau_{kick}$. 
Based on past experience, ELMs in JET are not expected to be triggered
with kicks of less than 10Wb (10 Volt seconds), so it seems safe to
assume that a kick cannot modify plasma stability until it has
produced at least 5Wb. 
After $\sim$1.3ms the kick's constant voltage produces a perturbation
to the radial magnetic field that grows approximately linearly with
time, at least for kick durations of less than 4-5ms or so. 
So for a kick with voltage $V_{kick}$ it seems reasonable to define a
minimum time $\tau_m$ for which $(\tau_m - 1.3)V_{kick}=5$,
i.e. $\tau_m=5/V_{kick} +1.3$, below which the kick has negligible
influence. 
If we take $2 \tau_{kick}$ as an upper time limit over which the kick
significantly perturbs the plasma, then $\Delta \tau = 2\tau_{kick} -
5/V_{kick} -1.3$. 
JET plasma pulse 83440 was based on the plasma in pulse
82630 over the period 18.5-20s, that was extended in 83440 to allow
kicks to be tested over an extended pulse duration. 
During the 18.5-20s period in 82630 there were $M=24$ ELMs, with an
average period $\bar{t}=0.059$ and standard deviation of $\sigma =
0.011$.  
This gives $\sigma/\bar{t}\sqrt{M} = 0.038$. 
Within the pulse 83440 the kick amplitude and duration are varied, as
shown in figure \ref{Webers}, with results as in figure
\ref{KickDetect},  and as recorded in table \ref{table1}.  
The different kick durations lead to different values of $\Delta \tau$
that are also recorded in table \ref{table1}, along with the number of kicks
in each set $N_K$, the number of ELMs that occur in the time
interval $\tau_m$ to $\tau_m+\Delta \tau$, and the values estimated
from that data for $P_{\bar{K}}$, $P_{\Delta \tau}$, $P(K)$, and their
errors. 
The quoted error for $P(K)$ is the largest of the upper and lower
errors as calculated using the estimates for $P_{\Delta \tau}$ and
$P_{\bar{K}}$. 
The Haldane prior biases the results towards kicks being
one of either successful or unsuccessful, and consequently because
estimates for $P(K)$ are all greater than $0.5$, the Haldane prior
gives higher kick-triggering probabilities than the uniform prior. 
Intuitively, the author feels more comfortable with the more
conservative, lower estimates for $P(K)$ that are obtained with a
uniform prior, although most present models for ELM triggering favour
a clear threshold for an ELM being triggered, or not. 
Independent of the choice of prior, from a physical point of view the
results are surprising, because 
they indicate that kicks with quite small amplitude can still trigger
ELMs if their duration is long enough. 
This has never previously been demonstrated, and can only be claimed
with confidence because of the analysis presented here. 
When applied to larger data sets, a similar analysis will allow the
quantitative assessment of how kick-triggering of ELMs depends on the
kick properties such as its amplitude, duration, total Webers (Volt
seconds), and maximum magnetic field perturbation. 
This will help us to understand how kicks trigger ELMs, and to
optimise the kick properties for triggering ELMs with the
``softest'' possible kick with lowest voltage for example. 

\begin{table}
\begin{center}
\normalsize
\begin{tabular}{|p{2.0cm}|p{1.4cm}|p{1.4cm}|p{1.4cm}|p{1.4cm}|p{1.4cm}|p{1.4cm}|p{1.4cm}|p{1.4cm}|p{1.4cm}|} 
\hline
\multicolumn{9}{|c|}{\bf Kick Properties of JET pulse 83440}
\\
\hline
$V_{kick}$ (kV) & 12 & 9 & 6 & 9 & 6 & 3 & 6 & 3  
\\
\hline
$\tau_{kick}$ (ms) & 2.5 & 2.5 & 2.5 & 3.3 & 3.3 & 3.3 & 4.5 & 4.5  
\\
\hline
$\Delta \tau$ (ms) & 3.3 & 3.1 & 2.9 & 4.7 & 4.5 & 3.6 & 6.9 & 6.0  
\\
\hline
$P_{\bar{K}}$  & 0.056 $\pm$ 0.002 & 0.053 $\pm$ 0.002 & 0.049 $\pm$ 0.002 &
  0.080 $\pm$ 0.003 & 0.076 $\pm$ 0.003 & 0.062 $\pm$ 0.002 & 0.116 $\pm$
  0.004 & 0.102 $\pm$ 0.004   
\\
\hline
$N_K$  & 15 & 15 & 15 & 14 & 14 & 14 & 14 & 14  
\\
\hline
$N$  & 14 & 13 & 9 & 14 & 14 & 9 & 14 & 11  
\\
\hline
\multicolumn{9}{|c|}{With uniform prior for $P(p)$}
\\
\hline
$P_{\Delta \tau}$  & 0.882 $\pm$ 0.076 & 0.824 $\pm$ 0.090 & 0.588
$\pm$ 0.116 & 0.938 $\pm$ 0.059 & 0.938 $\pm$ 0.059 & 0.625 $\pm$
0.117 & 0.938 $\pm$ 0.059 & 0.750 $\pm$ 0.105  
\\
\hline
$P(K)$  & 0.88 $\pm$ 0.09 & 0.81 $\pm$ 0.10 & 0.57 $\pm$ 0.12 & 0.93 
$\pm$ 0.07 & 0.93 $\pm$ 0.07 & 
0.60 $\pm$ 0.13 & 0.93 $\pm$ 0.07 &   0.72 $\pm$ 0.12
\\
\hline
\multicolumn{9}{|c|}{With Haldane prior $P(p)=1/p(1-p)$}
\\
\hline
$P_{\Delta \tau}$  & 0.933 $\pm$ 0.062 & 0.867 $\pm$ 0.085 & 0.600
$\pm$ 0.122 & 1.000 $\pm$ 0.000 & 1.000 $\pm$ 0.000 & 0.643 $\pm$
0.124 & 1.000 $\pm$ 0.000 & 0.786 $\pm$ 0.106  
\\
\hline
$P(K)$  & 0.93 $\pm$ 0.06 & 0.86 $\pm$ 0.09 & 0.58 $\pm$ 0.12 & 1.0
$\pm$ 0.00 & 1.00 $\pm$ 0.00 & 
0.62 $\pm$ 0.12 & 1.00 $\pm$ 0.00 &   0.76 $\pm$ 0.11
\\
\hline
\end{tabular}
\end{center}
\caption{
  The table summarises the kick properties and successes for
  JET pulse 83440, recording the voltage $V_{kick}$ and duration
  $\tau_{kick}$ of the kick. 
  $V_{kick}$ and $\tau_{kick}$ determine the duration $\Delta \tau$
  over which the kick influences the plasma, allowing an estimate for
  $P_{\bar{K}}$ to be calculated. 
  The number of kicks and the number of ELMs within a time interval  
  $\Delta \tau$ allows an estimate for $P_{\Delta \tau}$ to be
  calculated, and subsequently also for $P(K)$ the probability of the
  kick triggering an ELM. 
  Estimates are calculated with a uniform prior $P(p)$, and with the
  Haldane prior $P(p)=1/p(1-p)$.   
\label{table1}
}
\end{table}

\section{What can we learn about ELMs, and how to mitigate them?}  

The previous section mentioned some of the
opportunities that the techniques in this paper offer, the next few
paragraphs discuss some of the possibilities more fully. 
 
\subsection{Is ELM occurrence a deterministic process? Or is it 
  better regarded as statistical?} 

Even steady-state tokamak plasmas are turbulent, and
have continuously changing properties that are often best described in
a statistical way. 
In contrast, ideal MHD models for ELM instabilities have ELMs being
triggered after a specific threshold is exceeded in edge pressure
gradient or current  \cite{WebsterRev,Groebner}   
\footnote{The limit on edge current is a bit ambiguous, it depends
  sensitively  on how the edge-plasma boundary is numerically
  approximated. In the limit of the boundary asymptoting to a
  separatrix with one or more X-points, the peeling mode is stabilised
  by the strong magnetic shear near the X-point
  \cite{WebsterRev,Huysmanns,WebsterGimblettPRL}.  
  However, for an arbitrarily close approximation to the X-point,
  there are always unstable peeling modes, albeit with arbitrarily
  high mode numbers \cite{Laval, WebsterGimblettPRL}. 
}. 
A question is, is it more helpful to regard ELMs as a deterministic or
statistical process?   
There is growing evidence that a statistical description of ELMs might
be required.    
For example, \cite{WebsterDendy2013} shows many examples where the
times between ELMs can be described by a Weibull probability
distribution. 
This is consistent with the likelihood of observing an ELM in the next
fraction of time, increasing with the time since the previous ELM, so
it has elements of both a statistical and a deterministic model.
More strikingly, in a set of 120 identical JET plasma pulses where the
ELM energies were determined from the estimated losses of plasma
thermal energy, it was found that the majority of ELM energies were
statistically equivalent and independent of both the time since the
previous ELM and the size of the previous ELM \cite{SJWebster}.  
This seems contrary to the usual picture of ELMs, where
we would expect the time to the next ELM to be determined by the time
for the plasma's edge properties to be restored to their pre-ELM
values, and for this to be longer following a larger ELM. 
We might also have expected the ELM size to increase with time since
the previous ELM, but beyond an initial recovery period of about 20
milliseconds the ELM energies were found to be independent of the time
since the previous ELM \cite{SJWebster}.  


In principle vertical kicks provide an opportunity to test these
pictures, by establishing whether there is a fixed threshold beyond
which ELMs will be triggered and below which they cannot, or whether
there is a smoothly continuous increase in the probability of
triggering an ELM as the kick size is increased.   
In practice this study was not previously possible, because although
it was clear when all ELMs are being triggered by kicks, when
triggering is partially successful it is not possible to determine
whether any particular ELM has been triggered or is naturally
occurring, and there was no method to calculate the probability of ELMs
having been triggered.     
This situation has now changed. 
Consider the example described in Table 1, of a 2T 2MA JET pulse in an
otherwise steady-state H-mode with 10.2MW of neutral beam heating and
7.5$\times$10$^{22}$s$^{-1}$ of Deuterium gas fuelling. 
As the kick duration is reduced, with the kick voltage kept constant, the
probability of a kick triggering an ELM falls gradually. 
For example, with a uniform prior for $P(p)$: $V_{kick}$=9 and $P(K)$
falls from 0.93 ($\tau_{kick}$=3.3ms) to 0.81 ($\tau_{kick}$=2.5ms),
$V_{kick}$=6 and $P(K)$ falls from 0.93 ($\tau_{kick}$=3.3ms) to 0.57
($\tau_{kick}$=2.5ms), $V_{kick}$=3 and $P(K)$ falls from 0.72
($\tau_{kick}$=4.5ms) to 0.60 ($\tau_{kick}$=3.3ms); in none of these
cases does $P(K)$ drop sharply toward zero. 
Similar conclusions are reached if we use the Haldane prior.   
Therefore for this case it appears that there is not a sharp 
threshold between whether an ELM is successfully triggered or not, but
a gradual change in the probability of a kick being successful. 
Systematic experiments should be able to determine whether
kick-triggering of ELMs should be regarded as a statistical process,
and the circumstances for which this is the case, or not.    

\subsection{What are the thresholds for kicks to trigger ELMs?}

It would be of practical value to know what size and type of kick is
required to ensure a high probability of triggering ELMs.  
Table \ref{table1} suggests that kick-triggering of ELMs depends more
on the amplitude of the perturbation to the plasma, than the rate of
change of the perturbation.  
Because of inductance, and the large voltage of a kick, and the short
time-period over which the kick occurs, we might expect the rate of
change of the current in the vertical control system to be
approximately proportional to the voltage that is applied.  
Therefore if a sufficiently rapid rate of change in the magnetic
field's perturbation were required to trigger an ELM, then we would
expect kick triggering to become less effective when the voltage of
the kick is reduced from 12kV to 3kV, for which the rate of change of
the current would be reduced by roughly 75\%.  
However what is found is that provided the duration of the kick is
increased sufficiently, then the kicks can still trigger ELMs.  

\vspace{0.5cm}
\begin{figure}[htbp!]
\begin{center}
\includegraphics[width=11cm,height=11cm]{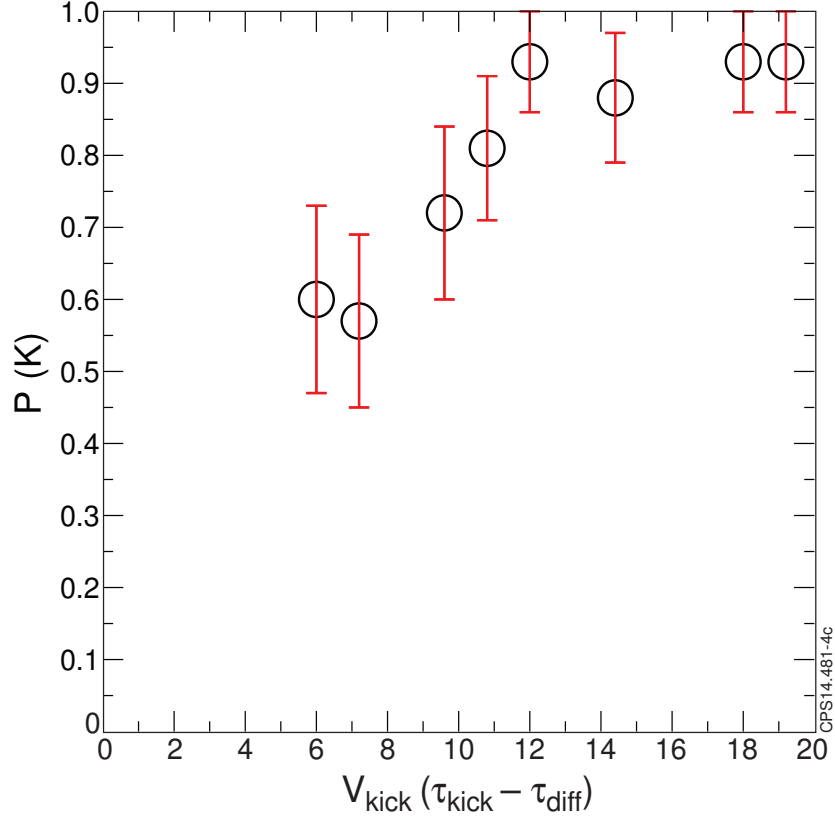}
\vspace{.0cm}
\vspace{0.8cm}
\caption{\label{Thresholds} 
Consider the results from table \ref{table1} with a constant prior,
for the 2T, 2MA, JET plasma 83440, that had 10.2MW of neutral beam 
heating and 7.5$\times$10$^{22}$s$^{-1}$ Deuterium gas fuelling. 
The probability of a kick triggering an ELM is plotted (vertical
axis), versus the product of kick duration ($\tau_{kick}$) minus the
1.3ms JET vacuum-vessel diffusion time ($\tau_{diff}$), and the kick
voltage ($V_{kick}$).  
The error bars are the standard deviations of $P(K)$ that are given in
table \ref{table1}.
When $V_{kick}(\tau_{kick}-\tau_{diff})$ exceeds 
12Wb (Volt seconds), the probability of a kick triggering an 
ELM appears to reach a maximum that is close to 1.  
(Higher values of $P(K)$ are only possible with a larger number of
kicks to improve the statistics, for which case the error bars will
also be reduced.) 
}
\end{center}
\end{figure}

We know that for JET there is a delay $\tau_{diff}$ of order 1.3ms for
the magnetic field to diffuse through the vacuum vessel, but otherwise
we might expect the maximum in the perturbation due to a kick to be
proportional to 
the voltage multiplied by the length of the kick. 
If we subtract off the $\tau_{diff}=$1.3ms delay from the kick's
duration (there will continue to be a small increase in the plasma
perturbation for a short period after the kick stops or reverses, but
we will neglect this), then multiply it by the 
voltage of the kick, this should be approximately proportional to the
maximum magnetic perturbation due to a kick. 
Plotting this against the probability of the kick being successful
(figure \ref{Thresholds}), it appears that the threshold for kick
success in this particular JET plasma, is a perturbation (to the plasma)
of order 12 Wb.  
In other words,for this plasma it appears that if the product of kick
duration minus the 1.3ms diffusion time,  and voltage
(i.e. $V_{kick}(\tau_{kick}-\tau_{diff})$), exceeds 12 Wb, then the
kicks were almost 100\% successful.    
Clearly this estimate needs refining with improved statistics
(i.e. more experimental data), and a more accurate modelling of the
peak magnetic perturbation to the plasma (by modelling how the
magnetic field perturbation diffuses through the vessel), but both of 
these are possible. 
This analysis can be repeated for different pulse types to obtain 
the requirements for triggering ELMs; or alternately 
to determine the maximum perturbations a control system can 
apply while still ensuring a low risk of triggering an ELM.   

\subsection{What triggers an ELM?}

Once the thresholds for which ELMs are triggered are known, then
modelling can determine the values of physical quantities such as the 
diffusing magnetic fields and currents, when those thresholds are
approached and passed. 
By determining which processes are present, absent, increasing, or
decreasing, hopefully we will be able to establish the 
mechanisms by which ELMs are being induced.  
For example, the rate of increase of the kick will 
determine how edge-localised any kick-induced currents will be. 
A short but rapidly increasing kick would be expected to produce
strong but very localised current perturbations to the plasma's edge,
but a longer and weaker kick would allow a weaker current perturbation
to diffuse further into the plasma.  
Peeling and peeling-ballooning modes
\cite{WebsterGimblettPRL,Laval,Lortz,Hegna,Connor}  
are widely recognised candidates for driving edge instabilities
\cite{WebsterRev,Groebner}, and are driven increasingly unstable by
increasingly strong currents at the plasma's edge. 
Therefore they might be expected to be more strongly destabilised by a
strong radially edge-localised current perturbation, as would be
greatest with the most rapidly increasing kicks, and smallest with a
longer and weaker kick.   
However for the example analysed here, the product of kick size and
duration was found to be more important than the rate at which the
kick was produced.  
Thorough modelling of these processes is probably required to
understand with certainty whether the peeling mode is involved in the
kick-triggering of ELMs, but the initial evidence just discussed
suggests that another mechanism might be important.   
A detailed modelling and discussion of the processes that occur during
a kick is beyond the scope of this paper, but there is enough evidence
presented here to indicate that a systematic experimental
investigation using the analysis tools in this paper can help  
clarify how a kick triggers an ELM; and possibly how ELMs are
triggered more generally.  

\section{Conclusions}

We have considered the generic problem of assessing whether an
otherwise quasi-random event such as an edge-localised plasma
instability (an ``ELM''), has been triggered by an external influence.   
The specific problem of assessing the success of experiments to
deliberately trigger ELMs by rapid vertical displacements of the
plasma has been analysed in detail, leading to a simple set of
rigorously derived formula to estimate the kick-triggering probability
and its accuracy.  
The advantage of the method is its rigour. 
For example, we can now assert with confidence that vertical kicks
with a voltage of only 3kV were in some cases triggering ELMs. 
This was unexpected, and has never previously been demonstrated. 
The method allows the success of kicks to be quantitatively assessed
with unprecedented accuracy, allowing systematic studies to
quantitatively determine how kick-triggering success depends on
properties such as their duration, amplitude, and the peak magnetic
field perturbation it produces. 
In the example considered (figure \ref{Thresholds}), the
probability of a kick being successful increased approximately linearly with
$V_{kick}(\tau_{kick}-\tau_{diff})$ until reaching a maximum of
approximately 1 at approximately 12Wb ($V_{kick}$ is the vertical
control system voltage, $\tau_{kick}$ is the duration of this step in
voltage, $\tau_{diff}$ is JET's vacuum vessel diffusion time).  
More generally, it is likely that analogous calculations and direct
adaptations of the method presented here can be used to help
distinguish cause from accidental correlation of quasi-random events
in other circumstances.

\vspace{0.5cm}
\begin{acknowledgments}
{\bf Acknowledgements:} 
AJW thanks Steve Fitzgerald for interesting past discussions about path
integral calculations that proved useful in this paper, and both Greg
Colyer and Richard Kemp for helpful comments and discussions about
different choices of prior $P(p)$.  
The kick-triggering experiments discussed in this paper were possible
due to the help and scientific co-ordination of Thomas Eich, Rory
Scannel, Peter Lang, Peter Lomas, and Fernanda Rimini. 
This work was supported by EURATOM and carried out within the
framework of the European Fusion Development Agreement. 
To obtain further information on the data and models underlying this
paper please contact PublicationsManager@ccfe.ac.uk. 
The views and opinions expressed herein do not necessarily reflect
those of the European Commission.   
\end{acknowledgments}

\newpage

\appendix

{\bf \large Appendix}

\section{The sum and product rules}

The basic sum and product rules of Bayesian probability theory
\cite{Jaynes} can be used to derive the key formulas used in this
paper. 
Firstly consider Eq. (\ref{e3}), noting that,
\begin{equation}\label{a1}
\begin{array}{ll}
P(A,B|C) + P(A,\bar{B}|C) &= \left[
P(B|A,C) + P(\bar{B}|A,C) 
\right] P(A|C) 
\\
&= P(A|C) 
\end{array}
\end{equation}
where the 1st line uses the basic product rule Eq. (\ref{e1}), and the
1st-2nd line uses the basic sum rule Eq. (\ref{e2}). 
Next using the product rule to expand $P(A,B|C)=P(A|B,C)P(B|C)$ and 
$P(A,\bar{B}|C)=P(A|\bar{B},C)P(\bar{B}|C)$, then (\ref{a1}) expands
to gives, 
\begin{equation}
P(A|C) = P(A|B,C)P(B|C) + P(A|\bar{B},C)P(\bar{B}|C) 
\end{equation}
which is Eq. (\ref{e3}). 
If we replace $A$ with $t$, $B$ with $K$, and $C$ with $\tau_m$, then
we have, 
\begin{equation}
P(t|\tau_m) = P(t|K,\tau_m)P(K|\tau_m) +
P(t|\bar{K},\tau_m)P(\bar{K}|\tau_m)  
\end{equation}
which is Eq. \ref{p0}, and gives $P(t|\tau_m)$ the probability density
of observing 
an ELM at time $t$ after a kick at time $\tau_m$, in
terms of $P(t|K,\tau_m)$ the 
probability density of observing an ELM at time $t$ after a kick at
time $\tau_m$ successfully triggers an ELM, 
and $P(t|\bar{K},\tau_m)$ the  probability density of
observing an ELM at time $t$ after a kick at time 
$\tau_m$ fails to trigger an ELM. 
The quantity $P(K|\tau_m)$ is the probability of a kick being
successful, given that the kick was at time $\tau_m$. 
As noted in the main text, the sum rule requires that
$P(\bar{K}|\tau_m)=1-P(K,\tau_m)$.

\section{Correlations between ELMs and kicks}\label{corr}

To assess how much correlations between the ELMs' natural frequency
and the kick frequency have the potential to influence the estimate,
we consider a situation where an ELM has naturally occurred within the
time window $\tau_m$ to $\tau_m + \Delta \tau$ that is usually
associated with kick-triggered ELMs, and where the average ELM waiting
time $\bar{t}$ equals the time between kicks $\tau_m-\tau_{m-1} $. 
Assuming the distribution of ELM waiting times is reasonably
approximated by a Gaussian distribution with $P(t)=(1/\sqrt{2\pi
  \sigma^2}) \exp\{ -(t-\bar{t})^2/2\sigma^2\}$, that can often be the
case for type I ELMs \cite{WebsterDendy2013}, then the
probability $p$ of an ELM between $\bar{t}$ to $\bar{t}+\Delta \tau$
is, 
\begin{equation}
p \simeq \Delta \tau
P(\bar{t}) = \frac{1}{\sqrt{2\pi}} \frac{\Delta \tau}{\sigma} 
\end{equation}
The probability of a sequence of $m$ ELMs within: $\bar{t}$ to
$(\bar{t} + \Delta \tau)$, $2\bar{t}$ to $(2\bar{t}+\Delta \tau)$, ... ,
$m\bar{t}$ to $(m\bar{t}+\Delta \tau)$, is then simply $p^m$, giving  
the number of correlated ELMs that we would expect to
observe as, 
\begin{equation}
\langle m \rangle = 
\frac{\sum_{m=1}^{\infty} m p^m}{\sum_{m=1}^{\infty} p^m} = 
\frac{p \frac{\partial}{\partial p}
\sum_{m=1}^{\infty} p^m
}{\sum_{m=1}^{\infty} p^m} 
= \frac{1}{1-p} 
\end{equation}
where we used $\sum_{m=1}^{\infty} p^m = p/(1-p)$ to
evaluate the sums. 
$\langle m \rangle$ tends to $1$ as $\Delta \tau/\sigma \rightarrow
0$, or infinity 
as $\Delta \tau/\sigma \rightarrow 1$. 
Most of the cases we are interested in have $\Delta \tau / \sigma \ll
1$.
For those cases, even if $\bar{t}=\tau_m-\tau_{m-1}$, then the number of
ELMs to be incorrectly counted as ``triggered'' by the kicks will be
small, with $\langle m \rangle <2$ for  $\Delta \tau / \sigma <0.5$,
and of order $1$ for $\Delta \tau / \sigma \ll 1$.  
In most cases, $\bar{t}$ will be substantially different to
$\tau_m-\tau_{m-1}$, and the potential influence of correlations can
be neglected entirely. 


\begin{thebibliography}{99}

\bibitem{Wesson} J. Wesson {\sl Tokamaks} (Oxford University Press, Oxford,
  1997).  

\bibitem{Zohm} H. Zohm, Plasma Physics and Controlled Fusion {\bf 38},
  105, (1996). 

\bibitem{Degeling} A.W. Degeling, Y.R. Martin, J.B. Lister,
  L. Villard, A.N. Dokouka, V.E. Lukash, R.R. Khayrutdinov, Plasma
  Phys. Control. Fusion, {\bf 45}, 1637, (2003). 

\bibitem{Lang} P.T. Lang, A.W. Degeling, J.B. Lister, Y.R. Martin,
  P.J. Mc Carthy, A.C.C. Sips, W. Suttrop, G.D. Conway, L. Fattorini,
  O. Gruber, L.D. Horton, A. Herrmann, M.E. Manso, M. Maraschek,
  V. Mertens, A. Mück, W. Schneider, C. Sihler, W. Treutterer, H. Zohm
  and ASDEX Upgrade Team,  
  Plasma Phys. Control. Fusion, {\bf 46}, L31, (2004). 

\bibitem{Liang} Y. Liang, Fusion Science and Technology {\bf 59}, 586,
  (2011). 

\bibitem{LangPacing} P.T. Lang, A. Loarte, G. Saibene, L.R. Baylor,
  M. Becoulet, M. Cavinato, S. Clement-Lorenzo, E. Daly, T.E. Evans,
  M.E. Fenstermacher, Y. Gribov, L.D. Horton, C. Lowry, Y. Martin,
  O. Neubauer, N. Oyama, M.J. Schaffer, D. Stork, W. Suttrop,
  P. Thomas, M. Tran, H.R. Wilson, A. Kavin, O. Schmitz 
  Nucl. Fusion {\bf 53}, 043004, (2013). 


\bibitem{KickRef} F.G. Rimini, F. Crisanti, R. Albanese,
  G. Ambrosino,M. Ariola,G. Artaserse, T. Bellizio, V. Coccorese,
  G. De Tommasi, P. De Vries, P. J. Lomas, F. Maviglia, A. Neto,
  I. Nunes, A. Pironti, G. Ramogida,  F. Sartori, S.R. Shaw,
  M. Tsalas, R. Vitelli, L. Zabeo, JET EFDA Contributors, 
  Engineering and Design, {\bf 86}, 539-543, (2011). 


\bibitem{Jaynes}
E.T. Jaynes ``Probability Theory the Logic of Science'', Cambridge
University Press, (2003). 

\bibitem{WebsterDendy2013} 
A.J. Webster and R.O. Dendy, Phys. Rev. Lett. {\bf 110}, 155004,
(2013).  

\bibitem{Sivia} D.S. Sivia ``Data Analysis a Bayesian Tutorial''
  Oxford University Press, Great Clarendon Street, Oxford, (1996). 

\bibitem{Schulman} L.S. Schulman ``Techniques and Applications of Path
  Integration'' John Wilely \& Sons, Inc. (1981). 

\bibitem{Brez}  S. Brezinsek, T. Loarer, V. Phillips, H.G. Esser,
  S. Gr\"unhagen, R. Smith, R. Felton, J. Banks, P. Belo, A. Boboc,
  J. Bucalossi, M. Clever, J.W. Coenen, I. Coffey, D. Douai,
  M. Freisinger, D. Frigione, M. Groth, A. Huber, J. Hobirk,
  S. Jachmich, S. Knipe, K. Krieger, U. Kruezi, G.F. Matthews,
  A.G. Meigs, F. Nave, I. Nunes, R. Neu, J. Roth, M.F. Stamp,
  S. Vartagnian, U. Samm, and JET EFDA contributors,   
  Nuclear Fusion, 53, 083023, (2013).

\bibitem{WebsterRev} A.J. Webster, Nucl. Fusion, 52, 114023, (2012). 

\bibitem{Groebner} R.J. Groebner, C.S. Chang, J.W. Hughes, R. Maingi,
  P.B. Snyder, 
X.Q. Xu, J.A. Boedo, D.P. Boyle, J.D. Callen, J.M. Canik, I. Cziegler,
E.M. Davis, A. Diallo, P.H. Diamond, J.D. Elder, D.P. Eldon,
D.R. Ernst, D.P. Fulton, M. Landreman, A.W. Leonard, J.D. Lore,
T.H. Osborne, A.Y. Pankin, S.E. Parker, T.L. Rhodes, S.P. Smith,
A.C. Sontag, W.M. Stacey, J. Walk, W. Wan, E.H.-J. Wang, J.G. Watkins,
A.E. White, D.G. Whyte, Z. Yan, E.A. Belli, B.D. Bray, J. Candy,
R.M. Churchill, T.M. Deterly, E.J. Doyle1, M.E. Fenstermacher,
N.M. Ferraro, A.E. Hubbard, I. Joseph, J.E. Kinsey, B. LaBombard,
C.J. Lasnier, Z. Lin, B.L. Lipschultz, C. Liu, Y. Ma, G.R. McKee,
D.M. Ponce, J.C. Rost, L. Schmitz, G.M. Staebler, L.E. Sugiyama,
J.L. Terry, M.V. Umansky, R.E. Waltz, S.M. Wolfe, L. Zeng and
S.J. Zweben, 
Nucl. Fusion, {\bf 53}, 093024, (2013). 


\bibitem{SJWebster} A.J. Webster, S.J. Webster, and JET-EFDA
  Contributors, arXiv:1311.1942, Accepted for publication in Physics
  of Plasmas, (2014).  


\bibitem{Laval} G. Laval, R. Pellat, J.S. Soule, Phys. Fluids, {\bf
    17}, 835, (1974). 

\bibitem{Lortz} D. Lortz, Nucl. Fusion {\bf 15}, 49, (1975).

\bibitem{WebsterGimblettPRL} A.J. Webster, C.G. Gimblett,
  Phys. Rev. Lett. {\bf 102}, 035003, (2009). 


\bibitem{Hegna} C.C. Hegna, J.W. Connor, R.J. Hastie, H.R. Wilson,
  Phys. Plasmas, {\bf 3}, 584, (1996). 

\bibitem{Connor} J.W. Connor, R.J. Hastie, H.R. Wilson, R.L. Miller,
  Phys. Plasmas, {\bf 5}, 2687, (1998). 

\bibitem{Huysmanns} G.T.A. Huysmanns, Plasma Phys. Controlled Fusion,
  {\bf 47}, 2107, (2005). 


\end{thebibliography}
\end{document}